\documentclass[12pt]{revtex4}
\usepackage{graphicx}
\usepackage{dcolumn}
\usepackage{bm}

\newcommand{\mb}[1]{ \mbox{\boldmath$#1$} }
\newcommand{\eps}{\varepsilon}
\newcommand{\ds}{\displaystyle}
\newcommand{\beq}{\begin{eqnarray}}
\newcommand{\eeq}{\end{eqnarray}}
\newcommand{\beqq}{\begin{eqnarray*}}
\newcommand{\eeqq}{\end{eqnarray*}}
\newcommand{\p}{\partial}

\newcommand{\x}{\mbox{\boldmath$x$}}
\newcommand{\y}{\mbox{\boldmath$y$}}
\newcommand{\z}{\mbox{\boldmath$z$}}
\font\bb=msbm10 at 12pt
\def\rR{\hbox{\bb R}}

\begin{document}

\title
{CONTINUOUS OBSERVATIONS AND THE WAVE FUNCTION COLLAPSE}

\author{A. Marchewka}
 \email{avi.marchewka@gmail.com}
 \affiliation{Department of Electrical and Electronics Engineering, College of Judea and Samaria, Ariel, Israel.}
\author{Z. Schuss}
 \email{schuss@post.tau.ac.il}
\affiliation{Department of Mathematics, Tel-Aviv University,
Ramat-Aviv, 69978 Tel-Aviv, Israel}

\date{\today}

\begin{abstract}
We propose to modify the collapse axiom of quantum measurement theory by
replacing the instantaneous with a continuous collapse of the wave function in
finite time $\tau$. We apply it to coordinate  measurement of a free quantum
particle that is initially confined to a domain $D\subset\rR^d$ and is observed
continuously by illuminating $\rR^d-D$. The continuous collapse axiom (CCA)
defines the post-measurement wave function (PMWF)in $D$ after a negative measurement
as the solution of Schr\"odinger's equation at time $\tau$ with instantaneously
collapsed initial condition and homogeneous Dirichlet condition on the boundary
of $D$. The CCA applies to all cases that exhibit the Zeno effect. It rids
quantum mechanics of the unphysical artifacts caused by instantaneous collapse
and introduces no new artifacts.
\end{abstract}

\pacs{03.65 Ta, 02.30 Rz, 03.65.Xp}

\maketitle

\section{Introduction}\label{sec:intro}

Schr\"odinger's equation does not describe the results of measurements. Rather,
a separate wave function collapse axiom \cite{coll}, \cite{Khalili} is needed
to connect between the Schr\"odinger evolution of the wave function and the
possible results of laboratory measurements. According to this axiom, as
applied to a quantum particle's coordinate, a measurement collapses the wave
function instantaneously to one that vanishes on a subset of positive measure
in the Euclidean space of the coordinate \footnote{The Dirac $\delta$-function
should be understood in this context as a function that vanishes outside a
sufficiently small interval that is the resolution of the measuring device.}.
Because all the possible collapsed wave functions of a given measurement form a
subspace of $L^2(\rR^n)$, the collapse is referred to as a projection into this
subspace \cite{coll}. According to the collapse axiom, the post-measurement
wave function (PMWF) evolves from its collapsed form according to
Schr\"odinger's equation.

Consider, for example, an ideal coordinate measurement of a particle by
illuminating instantaneously the positive axis $\rR^+$ or a finite interval and
assume that no particle is observed there \cite{pos}. The collapse axiom
implies that the PMWF is truncated instantaneously to zero on the positive axis
and that it is renormalized on the negative axis. This implies that after this
measurement the wave function is discontinuous at the origin and that it
evolves from its truncated form according to the Schr\"odinger equation
\cite{Khalili}. It was shown, however \cite{PLA}, that the evolved PMWF has no
moments, no average momentum, and has infinite energy, which is unphysical,
because only finite energy is expended in this measurement. The same phenomenon
occurs if the result of the measurement is that the particle is in the positive
axis, without specifying its location there. Infinite measurable physical
quantities, such as moments, are not encountered in the physical world,
therefore infinities are incompatible with classical, relativistic, or quantum
physics. The problem of infinities permeates to the foundations of theoretical
physics (the Copenhagen interpretation \cite{chop} and weak causality
\cite{Heg1}).

Despite these difficulties, the results of the collapse axiom describe
faithfully laboratory quantum measurements, at least as a phenomenological and
operational theory. So far, unfortunately, Hamiltonian theories have failed to
reconcile the collapse axiom with Schr\"odinger's equation. Roughly speaking,
there are two main approaches for alleviating the problem, one is to modify the
collapse axiom, as we propose below, and the other is to modify the
Schr\"odinger equation as in e.g. \cite{Tempered}, \cite{pearl1}, \cite{deco}.

Our aim in this paper is to reconcile the collapse axiom for coordinate
measurement with Schr\"odinger's equation by giving up the assumption of
instantaneous truncation as the description of a coordinate measurement.
Rather, we assume that a single coordinate measurement is continuous and has
finite duration. Specifically, we propose to modify the collapse axiom for the
negative coordinate measurement by postulating that it lasts for a positive
time $\tau$, specific to each measurement apparatus, and the PMWF the
unmeasured domain is the solution of Schr\"odinger's equation at time $\tau$
with initial condition that is the truncated wave function and a zero Dirichlet
(totally reflecting) condition on the boundary of the domain.

The resulting PMWF is called the continuously collapsed wave function.
Although, to the best of our knowledge, our continuous collapse axiom,
similarly to the instantaneous collapse axiom, has no Hamiltonian realization,
we show below that, unlike the collapse axiom, it does not introduce artifacts
into single particle quantum theory. We introduce the finite duration of the
continuous coordinate measurement to reflect the fact that all physical
measurements require finite time and this time is the property of every
individual measurement apparatus \cite{deco}. The continuously collapsed wave
function has a finite first moment. We note that continuous collapse for
measurements that exhibit the Zeno effect  leaves the state unchanged, thus it
introduces no new phenomenology.

\section{Continuous observation of a Brownian particle}\label{sec:2}

We calculate the probability density function of a Brownian particle by
considering its intermittent observations. A free Brownian particle is
initially confined to a domain $D\subset\rR^d$ and is observed intermittently
by instantaneous illuminations of $\Omega=\rR^d-D$ at times $\Delta t$ apart.
Between observations the particle diffuses freely. It is our purpose to
evaluate the pdf $p(\x,t)$ of the particle in $D$ at observation times $\Delta
t,2\Delta t,\dots,N\Delta t$, given that it was not observed in $\Omega$. We
begin with evaluating the pdf in the first interval $[0,\Delta t]$. At time
$\Delta t$ the pdf in $\rR^d$ is given by
 \beq
p_1(\x_1,\Delta t)=(2\pi\Delta
t)^{-d/2}\int\limits_Dp_0(\x_0)\exp\left\{-\frac{|\x_1-\x_0|^2}{2\Delta
t}\right\}\,d\x_0.
 \eeq
After the instantaneous observation at time $\Delta t$ the pdf in $D$ is the
conditional density $\pi_1(\x_1,\Delta t)$, given that the particle is not in
$\Omega$. That is,
 \beq
\pi_1(\x_1,\Delta t)=\frac{p_1(\x_1,\Delta t)}{\ds\int\limits_Dp_1(\x_1,\Delta
t)\,d\x_1},
 \eeq
which is normalized in $D$. At time $2\Delta t$ the propagated density is
 \beq
p_2(\x_1,2\Delta t)=(2\pi\Delta t)^{-d/2}\int\limits_D\pi_1(\x_1,\Delta
t)\exp\left\{-\frac{|\x_2-\x_1|^2}{2\Delta t}\right\}\,d\x_1,
 \eeq
the conditional density is
 \beq
\pi_2(\x_2,2\Delta t)&=&\frac{p_2(\x_2,2\Delta
t)}{\ds\int_Dp_2(\x_2,2\Delta t)\,d\x_2}\nonumber\\
&=&\frac{(2\pi\Delta t)^{-d/2}\ds\int_D\pi_1(\x_1,\Delta
t)\exp\left\{-\frac{|\x_2-\x_1|^2}{2\Delta t}\right\}\,d\x_1}{(2\pi\Delta
t)^{-d/2}\ds\int_D\int_D\pi_1(\x_1,\Delta
t)\exp\left\{-\frac{|\x_2-\x_1|^2}{2\Delta
t}\right\}\,d\x_1\,d\x_2},\label{rec1}
 \eeq
and so on. The recursion for $\pi(\x,t)$ on the lattice $t=j\Delta t$
($j=1,2,\ldots$) is therefore
 \beq
\pi(\x,t+\Delta t)=\frac{(2\pi\Delta
t)^{-d/2}\ds\int_D\pi(\y,t)\exp\left\{-\frac{|\x-\y|^2}{2\Delta
t}\right\}\,d\y}{(2\pi\Delta t)^{-d/2}\ds\int_D\int_D\pi(\y,
t)\exp\left\{-\frac{|\y-\z|^2}{2\Delta t}\right\}\,d\y\,d\z}.\label{rec2}
 \eeq

The integral in the denominator of (\ref{rec2}) can be evaluated by the change
of variables $\y=\z+\mb{\xi}\sqrt{\Delta t}$ as
 \beq
&&(2\pi)^{-d/2}\ds\int\limits_D\int\limits_D\pi(\z+\mb{\xi}\sqrt{\Delta t},
t)\exp\left\{-\frac{|\mb{\xi}|^2}{2}\right\}\,d\y\,d\z\nonumber\\
&=&(2\pi)^{-d/2}\ds\int\limits_D\int\limits_D\left[\pi(\z,t)+\sqrt{\Delta
t}\mb{\xi}\cdot\nabla_{\x}\pi(\z,t)+\frac{\Delta
t}{2}\sum_{i,j}\xi_i\xi_j\pi_{x_ix_j}(\z,t)+O\left(|\mb{\xi}\sqrt{\Delta t}|^3\right)\right]\nonumber\\
&&\times
\exp\left\{-\frac{|\mb{\xi}|^2}{2}\right\}\,d\mb{\xi}\,d\z=\ds\int\limits_D\pi(\z,t)\,d\z+\frac{\Delta
t}{2}\oint\limits_{\p D}\frac{\p\pi(\z,t)}{\p n}\,dS_{\z}+o(\Delta t)\nonumber\\
&=&1+\Delta tJ(t)+o(\Delta t),\label{denom}
 \eeq
where $\mb{n}$ is the unit outer normal at the boundary and $J(t)$ is the total
absorption flux on the boundary, given by
 \beq
 J(t)=\frac12\oint\limits_{\p D}\frac{\p\pi(\z,t)}{\p n}\,dS.\label{Jt}
 \eeq

Expanding the left hand side and the integral in the numerator of equation
(\ref{rec2}), we obtain from (\ref{denom}) that for small $\Delta t$,
 \beq
 (1+\Delta tJ(t))[\pi(\x,t)+\Delta t\pi_t(\x,t)+o(\Delta
 t)]=\pi(\x,t)+\frac{\Delta t}{2}\Delta_{\x}\pi(\x,t)+o(\Delta
 t).
 \eeq
Hence, in the limit $\Delta t\to0$,
 \beq
 \pi_t(\x,t)=\frac{1}{2}\Delta_{\x}\pi(\x,t)-J(t)\pi(\x,t)\hspace{0.5em}\mbox{for}\ \x\in
 D.\label{pieq}
 \eeq
If $\x\in\p D$ in (\ref{rec2}), then in the limit $\Delta t\to0$ the Gaussian
integral in the numerator extends over half the space (see \cite{PRA},
\cite{DSP}), which leads to the absorbing boundary condition
 \beq
 \pi(\x,t)=0\hspace{0.5em}\mbox{for}\ \x\in\p D.\label{pibc}
 \eeq
The initial condition for $\pi(\x,t)$ is
 \beq
 \pi(\x,0)=p_0(\x)\hspace{0.5em}\mbox{for}\ \x\in D.\label{piic}
 \eeq

The solution of the nonlinear initial boundary value problem
(\ref{pieq})-(\ref{piic}) can be constructed in the form of the renormalized
density
 \beq
 \pi(\x,t)=\frac{p(\x,t)}{\ds\int_Dp(\y,t)\,d\y},\label{pip}
 \eeq
where $p(\x,t)$ is the solution of the Fokker-Planck equation in $D$ with
absorbing boundary conditions on $\p D$,
 \beq
 p_t(\x,t)&=&\frac12\Delta_{\x}p(\x,t)\hspace{0.5em}\mbox{for}\
 \x\in D,\,t>0\label{DE}\\
 p(\x,t)&=&0\hspace{0.5em}\mbox{for}\
 \x\in\p D,\,t>0\label{DBC}\\
 p(\x,0)&=&p_0(\x)\hspace{0.5em}\mbox{for}\
 \x\in D.\label{DIC}
 \eeq
Indeed, differentiating (\ref{pip}) once with respect to $t$  and twice with
respect to $\x$ gives
 \beqq
 \pi_t(\x,t)&=&\frac{p_t(\x,t)}{\ds\int_Dp(\y,t)\,d\y}-\frac{p(\x,t)\,\ds\int_Dp_t(\y,t)\,d\y
 }{\left(\ds\int_Dp(\y,t)\,d\y\right)^2}\\
&=&\frac{\frac12\Delta_{\x}p(\x,t)}{\ds\int_Dp(\y,t)\,d\y}-\frac{p(\x,t)
\frac12\ds\int_D\Delta_{\x}p(\y,t)\,d\y}{\left(\ds\int_Dp(\y,t)\,d\y\right)^2}\\
&=&\frac12\Delta\pi(\x,t)-J(t)\pi(\x,t),
 \eeqq
which is (\ref{pieq}). We have used the identity
 \beqq
J(t)=\frac{\frac12\ds\int_D\Delta_{\x}p(\y,t)\,d\y}{\ds\int_Dp(\y,t)\,d\y},
 \eeqq
which is (\ref{Jt}).

The boundary value (\ref{pieq})-(\ref{piic}) suggests the following Brownian
simulation of the continuous observation process. At each time step $\Delta t$
of the simulation returns the escaping particles to $D$ and distributes them
there according to the existing (empirical) density. This, in effect, amounts
to putting sources distributed in $D$ according to the instantaneous density
and the strength of each source is the total efflux on the boundary.

\section{Continuous coordinate measurement of a quantum particle}\label{s:IOQP}
We adopt the above procedure to the coordinate observation of a quantum
particle. The support of the freely propagating wave function is collapsed to
$D$ at times of negative observations. It follows that at observation times
 \beq \psi(\x,t+\Delta
t)=(2\pi i\Delta
t)^{-d/2}\ds\int\limits_D\psi(\y,t)\exp\left\{-\frac{|\x-\y|^2}{2i\Delta
t}\right\}\,d\y,\label{prop3}
 \eeq
as shown in \cite{PLA0}, \cite{PRA1}. In the limit $\Delta t\to0,\ N\Delta t\to
t$ the solution of equation (\ref{prop3}) converges to the solution of
Schr\"odinger's equation in $D$ with the totally reflecting boundary condition
$\psi(\x,t)=0$ for all $\x\in\p D$ \cite{PLA}. Therefore continuous
observations do not allow the particle to exit $D$. This is the Zeno paradox in
the sense that the wave function in the observed domain $\rR^+$ is frozen at 0
\cite{Friedman}, \cite{SchulmanZeno}, \cite{zenopo}.

However, if the wave function is renormalized after each observation, a
procedure analogous to that of Section \ref{sec:2} gives for the renormalized
wave function the recursion
 \beq \pi(\x,t+\Delta
t)=\frac{(2\pi i\Delta
t)^{-d/2}\ds\int_D\pi(\y,t)\exp\left\{-\frac{|\x-\y|^2}{2i\Delta
t}\right\}\,d\y}{\left\{\ds\int_D\left|(2\pi i\Delta t)^{-d/2}\int_D\pi(\y,
t)\exp\left\{-\frac{|\y-\z|^2}{2i\Delta
t}\right\}\,d\y\right|^2\,d\z\right\}^{1/2}}.\label{rec3}
 \eeq
If $\x\in\p D$, then in the limit $\Delta t\to0$ we obtain, as in Section
\ref{sec:2}, that $\pi(\x,t)=0$.

To determine the differential equation that $\pi(\x,t)$ satisfies in $D$, we
expand the denominator as in (\ref{denom}). With the substitution
$\y=\z+\mb{\xi}\sqrt{\Delta t}$, the inner integral is
 \beq
&&(2\pi i\Delta t)^{-d/2}\int\limits_D\pi(\y,
t)\exp\left\{-\frac{|\y-\z|^2}{2i\Delta t}\right\}\,d\y\nonumber\\
&=&(2\pi i)^{-d/2}\ds\int\limits_{(D-z)/\sqrt{\Delta
t}}\pi(\z+\mb{\xi}\sqrt{\Delta t},
t)\exp\left\{-\frac{|\mb{\xi}|^2}{2i}\right\}\,d\mb{\xi}\nonumber\\
&=&(2\pi i)^{-d/2}\ds\int\limits_{(D-z)/\sqrt{\Delta
t}}\left[\pi(\z,t)+\sqrt{\Delta
t}\mb{\xi}\cdot\nabla_{\x}\pi(\z,t)+\frac{\Delta
t}{2}\sum_{i,j}\xi_i\xi_j\pi_{x_ix_j}(\z,t)+O\left(|\mb{\xi}\sqrt{\Delta t}|^3\right)\right]\nonumber\\
&&\times \exp\left\{-\frac{|\mb{\xi}|^2}{2i}\right\}\,d\mb{\xi}. \label{denomi}
 \eeq
The divergent integrals have to be summed by replacing
$\exp\left\{-|\mb{\xi}|^2/2i\right\}$ with
$\exp\left\{-|\mb{\xi}|^2(1+i\eps)/2i\right\}$ for a positive $\eps$ and taking
first the limit $\Delta t\to0$ and then $\eps\to0$.

In the one-dimensional case $D=[-a,a]$, so the first term gives
 \beqq
 &&(2\pi i)^{-1/2}\pi(z,t)\ds\int\limits_{(-a-z)/\sqrt{\Delta
t}}^{(a-z)/\sqrt{\Delta
t}}\exp\left\{-\frac{\xi^2(1+i\eps)}{2i}\right\}\,d\xi\\
&=&\pi(z,t)(1+o(\Delta t)),
 \eeqq
the second term gives
 \beqq
&&\frac{\sqrt{\Delta t}}{\sqrt{2\pi
i}}\pi_x(z,t)\ds\int\limits_{(-a-z)/\sqrt{\Delta t}}^{(a-z)/\sqrt{\Delta
t}}\xi\exp\left\{-\frac{\xi^2(1+i\eps)}{2i}\right\}\,d\xi=o(\Delta t),
 \eeqq
and the third term is
 \beqq
&&\frac{\Delta t}{2\sqrt{2\pi
i}}\pi_{xx}(z,t)\ds\int\limits_{(-a-z)/\sqrt{\Delta t}}^{(a-z)/\sqrt{\Delta
t}}\xi^2\exp\left\{-\frac{\xi^2(1+i\eps)}{2i}\right\}\,d\xi\\
&=&\frac{i\Delta
t}{2(1+i\eps)}\pi_{xx}(z,t)\left[1+o(1)\right]\hspace{0.5em}\mbox{for}\ \Delta
t\to0.
 \eeqq
Thus the only term in the denominator of (\ref{rec3}) that is $O(\Delta t)$ is
 \beqq
&&\frac{\Delta
t}{2}\int\limits_{-a}^a\left[\frac{i}{1+i\eps}\bar\pi(z,t)\pi_{xx}(z,t)-\frac{i}{1-i\eps}
\pi(z,t)\bar\pi_{xx}(z,t)\right]\,dz\\
&=&i\eps\Delta t\int\limits_{-a}^a\frac{|\pi_x(z,t)|^2}{1+\eps^2}\,dz.
 \eeqq
In higher dimensions this term is
 \beq
i\eps\Delta
t\int\limits_{D}\frac{|\nabla\pi(\z,t)|^2}{1+\eps^2}\,d\z(1+o(1)).\label{Sflux}
 \eeq
Now it follows from (\ref{rec3}) and (\ref{Sflux}) that
 \beq
\pi(\x,t+\Delta t)=\frac{(2\pi i\Delta
t)^{-d/2}\ds\int_D\pi(\y,t)\exp\left\{-\frac{|\x-\y|^2}{2i\Delta
t}\right\}\,d\y}{\left[1+i\eps\Delta
t\ds\int_{D}\frac{|\nabla\pi(\z,t)|^2}{1+\eps^2}\,d\z(1+o(1))\right]^{1/2}},\label{rec4}
 \eeq
which, as in (\ref{pieq}), gives the nonlinear Schr\"odinger equation
 \beq
\pi_t(\x,t)=\frac{i}{2}\Delta_{\x}\pi(\x,t)-J(t)\pi(\x,t),\label{piSE}
 \eeq
where
 \beq
J(t)=\frac{i\eps}{2}\int\limits_{D}\frac{|\nabla\pi(\z,t)|^2}{1+\eps^2}\,d\z\to0\hspace{0.5em}\mbox{as}\
\eps\to0.\label{epsJ}
 \eeq

As far as the question of the need for renormalization is concerned, similarly
to the case of diffusion (see Section \ref{sec:2}), the conditional probability
density of the quantum particle in $D$ after a negative observation is
$|\pi(\x,t+\Delta t)|^2$, as given in (\ref{rec3}). Therefore (\ref{rec3})
holds up to a pure phase factor, which can be assumed to be 1. It follows from
(\ref{piSE}) and (\ref{epsJ}) that the wave function of a quantum particle
under continuous observation is the solution of the Schr\"odinger equation in
$D$ with homogeneous Dirichlet boundary conditions, with or without
renormalization. Summa summarum, (\ref{piSE}) and (\ref{epsJ}) show that the
matter is mute.

Thus the wave function of a quantum particle under continuous negative
observations of the spatial coordinate for any period $\tau$ is the solution
$\pi(\x,\tau)$ of the Schr\"odinger equation in $D$ with truncated initial
conditions in $D$ and homogeneous Dirichlet boundary conditions on $\p D$. One
consequence of this observation is that if the wave function collapses in
$\rR^+$ at time $t=0$, it has to stay collapsed throughout the measurement
period $[0,\tau]$. Indeed, if it vanishes in the illuminated domain $\rR^+$ at
$t=0$, but does not vanish there at positive time $0<t\leq\tau$, then the wave
stays collapsed throughout the interval $[0,t)$, so according to
Schr\"odinger's equation, as described above, it vanishes on $\rR^+$ at time
$t$ as well and there can be no particle detected at this time. This fact is a
manifestation of the Zeno effect, when applied to coordinate measurement
\cite{pos}, \cite{SchulmanZeno}, \cite{zenopo}.

We call the solution of Schr\"odinger's equation with zero boundary conditions
the continuously collapsed wave function. The above discussion shows that
unlike the assertion of the collapse axiom, that after a single collapse in the
observed region the wave function in the unmeasured region $D$ stays intact,
continuous spatial coordinate observations for any time $\tau$ change the wave
function in the unmeasured domain, as described above, but freezes in the
measured region. More specifically, the continuously collapsed wave function in
$D$ becomes $\pi(\x,\tau)$.
\section{Discussion}

An experimental realization of continuous measurement is to place photographic
plates parallel to the direction of the light in the measured region $\rR^+$ so
that photons scattered from a particle that entered $\rR^+$ are absorbed by the
plates. The laboratory absorption times can be recorded to construct a time
histogram of measured particles. If at the end of the exposure period $\tau$
the photographic plate detects no particle, we have to conclude that throughout
this period there was no particle in the measured region and the wave function
vanished there. It follows that the wave function at time $\tau$ is the
solution of Schr\"odinger's equation with zero boundary conditions. Pursuing
the same logic, as mentioned above, if there is collapse of the wave function
at time $t=0$, the wave function stays collapsed throughout the measurement
period $0\leq t\leq\tau$. Indeed, if the time histogram shows a particle at
time $0<t\leq\tau$, then the wave function stayed collapsed throughout the
interval $[0,t)$, so according to Schr\"odinger's equation, as described in
Section \ref{s:IOQP}, it vanishes on $\rR^+$ at time $t$ as well and there can
be no particle detected at this time. This fact is a manifestation of the Zeno
effect, when applied to coordinate measurement.

As mentioned in the Introduction, we propose here a modification of the
collapse axiom for the negative coordinate measurement by postulating that it
lasts for a positive time $\tau$ and the PMWF in the unmeasured domain is the
solution of Schr\"odinger's equation at time $\tau$ with initial condition that
is the truncated wave function and a zero Dirichlet condition on the boundary
of the domain. It follows that the PMWF $\pi(x,\tau)$ is continuous on the
entire line and has a finite derivative at the boundary. It was shown in
\cite{PRA} that the solution of Schr\"odinger's equation on the entire line,
with a continuous initial condition and finite one-sided derivatives decays at
infinity as $|x|^{-2}$, so it has a finite first moment. Thus the continuous
collapse axiom alleviates the artifact, mentioned in the Introduction,
introduced by the instantaneous collapse. Note that continuous collapse for
measurements that exhibit the Zeno effect leaves the state unchanged, thus it
introduces no new phenomenology.

\begin{figure}
\resizebox{!}{6cm}{\includegraphics{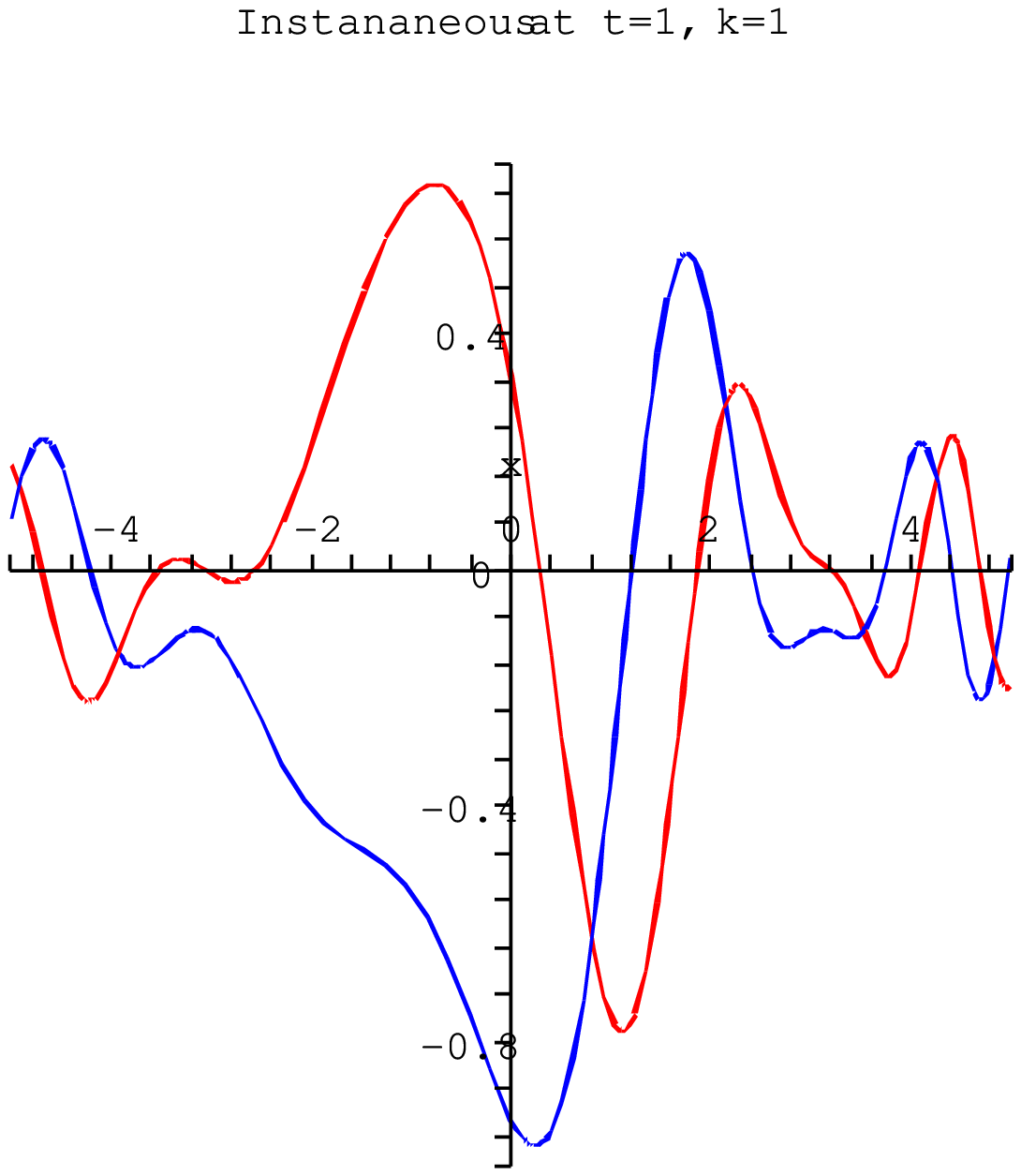}}\resizebox{!}{6cm}{\includegraphics{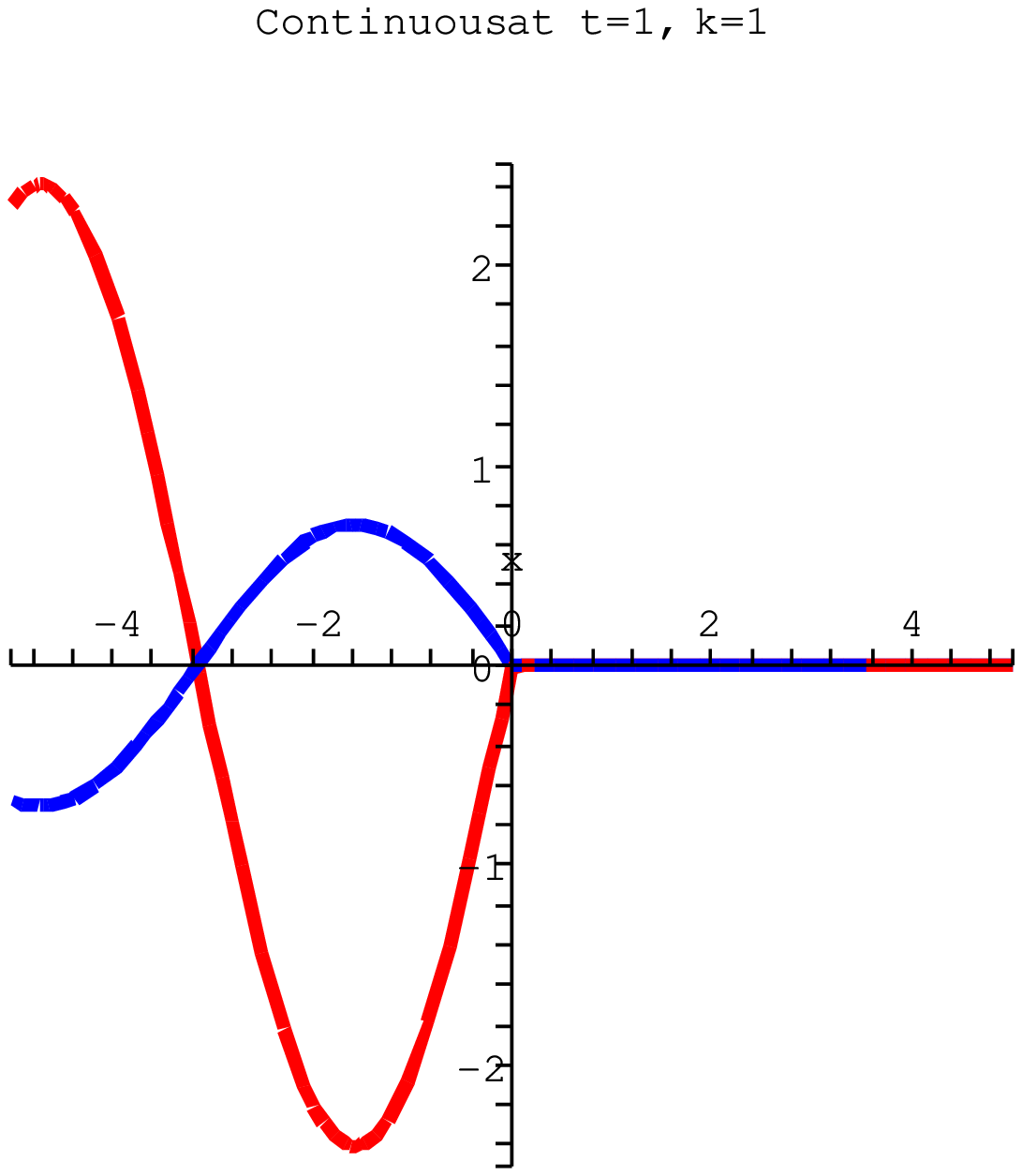}}
\caption{\small The real (red) and imaginary (blue) parts of $\varphi^I(x,1)$
(left) and of $\varphi^C(x,1)$ (right).} \label{f:exp-inst}
\end{figure}

There is, however, a difference between the experimental measurement of the
wave function at time $\tau$ in the instantaneous and continuous cases.
Consider the initial pre-measurement plane-wave (in normalized dimensionless
variables)
 \beq
 \psi(x)=e^{-ix},
 \eeq
which is measured at time $t=0$ a negative measurement on the positive axis. In
case of an instantaneous collapse, at time $t=\tau$ the truncated wave function
  \beq
 \varphi^I(x,0)=\Theta(-x)e^{-ix}\label{twf}
 \eeq
will have propagated into
 \beq
 \varphi^I(x,\tau)=\frac{1}{\sqrt{i\pi
 \tau}}\int\limits_{-\infty}^0e^{-iy}e^{-i(x-y)^2/2\tau}\,dy.
 \eeq
According to Section \ref{s:IOQP}, in the case of a continuous collapse that
ends at time $t=\tau$ the continuously collapsed wave function is the solution
of Schr\"odinger's equation in $\rR^-$ at time $\tau$ with zero boundary
condition at $x=0$ and initial condition (\ref{twf}) and it vanishes in
$\rR^+$. This solution is constructed by reflection in the origin as will have
propagated into
 \beq
 \varphi^C(x,\tau)=\frac{2i\Theta(-x)}{\sqrt{i\pi
 \tau}}\int\limits_{-\infty}^\infty\sin ky\,e^{-i(x-y)^2/2\tau}\,dy.
 \eeq
Figure \ref{f:exp-inst} shows the real (red) and imaginary (blue) parts of
$\varphi^I(x,1)$ (left) and $\varphi^C(x,1)$ (right). The two different
statistics should be observable.
Comparing the PMWF, we see that $\varphi^I(x,0)=\Theta(-x)$ while
$\varphi^C(x,\tau)$ is that given in Figure \. There should be also a
difference in the measurement of the average spatial coordinate in a given
interval after an instantaneous negative coordinate measurement and after a
negative continuous coordinate measurement. Specifically, in the former case,
the average coordinate in the interval $[x_1,x_2]$ at time $\Delta t$ after the
collapse is $O(\Delta t^{1/2}\log(x_2-x_1))$, whereas in the latter case it is
$O(\Delta t^{3/2}(x_1^{-2}-x_2^{-2}))$ \cite{PRA}, \cite{PRA1}.

The implication of the last observation in Section \ref{s:IOQP} is that if
intermittent absorption of a free quantum particle on the positive axis $\rR^+$
is defined as a process of instantaneous truncation of the wave function on
$\rR^+$ at times $\Delta t$ apart, then, in contrast to the case of
intermittent measurements, the wave function on the negative axis is not
renormalized. Thus, similarly to the case of measurement, intermittent
instantaneous absorption does not permit the particle to propagate into
$\rR^+$, resulting in no absorption at all.

The analogous situation in diffusion is different. Assume that intermittent
instantaneous absorption in $\rR^+$ is defined as turning on intermittently an
instantaneous infinite killing measure $k(x,t)$ in $\rR^+$, for example,
 \beqq
k(x,t)=\sum\limits_{k=1}^N\delta(t-k\Delta t)\Theta(x),
 \eeqq
where $N=t/\Delta t$ and $\Theta(x)$ is the Heaviside step function. Then,
according to the Feynman-Kac formula \cite{DSP}, the transition probability
density of the killed Brownian motion is the solution of the problem
 \beq
\frac{\p p_N(x,t)}{\p t}=\frac12\frac{\p^2p_N(x,t)}{\p
x^2}-\sum\limits_{k=1}^N\delta(t-k\Delta t)\Theta(x)p_N(x,t).
 \eeq
Integrating over space and time, we obtain
 \beq
\int\limits_{-\infty}^\infty p_N(x,t)\,dx-1=\frac{1}{\Delta
t}\int\limits_0^\infty \sum\limits_{k=1}^N p_N(x,t-k\Delta t)\Delta
t\,dx.\label{ints}
\eeq

\end{document}